\documentclass[10pt, conference]{IEEEtran}
\IEEEoverridecommandlockouts

\usepackage{algorithmic}
\usepackage[ruled,vlined]{algorithm2e}

\usepackage{cite}
\usepackage{amsmath,amssymb,amsfonts}
\usepackage{algorithmic}
\usepackage{graphicx}
\usepackage{textcomp}
\usepackage{xcolor}
\usepackage{tikz}
\usepackage{url}
\usepackage[hidelinks]{hyperref}
\usepackage[nameinlink]{cleveref}
\def\checkmark{\tikz\fill[scale=0.4](0,.35) -- (.25,0) -- (1,.7) -- (.25,.15) -- cycle;} 
\def\BibTeX{{\rm B\kern-.05em{\sc i\kern-.025em b}\kern-.08em
    T\kern-.1667em\lower.7ex\hbox{E}\kern-.125emX}}
\begin{document}

\newcommand\copyrighttext{%
  \footnotesize \textcopyright 2022 IEEE. Personal use of this material is permitted. Permission from IEEE must be obtained for all other uses, in any current or future media, including reprinting/republishing this material for advertising or promotional purposes, creating new collective works, for resale or redistribution to servers or lists, or reuse of any copyrighted component of this work in other works. Cite this article as follows: S. \v{S}pa\v{c}ek, P. Velan, P. \v{C}eleda, and D. Tovar\v{n}\'ak. \textit{HTTPS Event-Flow Correlation: Improving Situational Awareness in Encrypted Web Traffic}, NOMS 2022-2022 IEEE/IFIP Network Operations and Management Symposium, 2022, pp. 1-6, doi: \href{https://doi.org/10.1109/NOMS54207.2022.9789877}{10.1109/NOMS54207.2022.9789877}.}
\newcommand\copyrightnotice{%
\begin{tikzpicture}[remember picture,overlay]
\node[anchor=south,yshift=12pt] at (current page.south) {\fbox{\parbox{\dimexpr\textwidth-\fboxsep-\fboxrule\relax}{\copyrighttext}}};
\end{tikzpicture}%
}

\title{HTTPS Event-Flow Correlation: Improving Situational Awareness in Encrypted Web Traffic}

\author{\IEEEauthorblockN{Stanislav Špaček}
\IEEEauthorblockA{\textit{Institute of Computer Science} \\
\textit{and Faculty of Informatics} \\
\textit{Masaryk University}\\
Brno, Czech Republic \\
spaceks@ics.muni.cz}
\and
\IEEEauthorblockN{Petr Velan}
\IEEEauthorblockA{\textit{Institute of Computer Science} \\
\textit{Masaryk University}\\
Brno, Czech Republic \\
velan@ics.muni.cz}
\and
\IEEEauthorblockN{Pavel Čeleda}
\IEEEauthorblockA{\textit{Institute of Computer Science} \\
\textit{Masaryk University}\\
Brno, Czech Republic \\
celeda@ics.muni.cz}
\and
\IEEEauthorblockN{Daniel Tovarňák}
\IEEEauthorblockA{\textit{Institute of Computer Science} \\
\textit{Masaryk University}\\
Brno, Czech Republic \\
tovarnak@ics.muni.cz}
}

\maketitle

\copyrightnotice

\begin{abstract}

Achieving situational awareness is a challenging process in current HTTPS-dominant web traffic.
In this paper, we propose a new approach to encrypted web traffic monitoring.
First, we design a method for correlating host-based and network monitoring data based on their common features and a correlation time-window.
Then we analyze the correlation results in detail to identify configurations of web servers and monitoring infrastructure that negatively affect the correlation.
We describe these properties and possible data preprocessing techniques to minimize their impact on correlation performance.
Furthermore, to test the correlation method's behavior in different web server setups and for recent encryption protocols, we modify it by adapting the correlation features to TLS 1.3 and QUIC.
Finally, we evaluate the correlation method on a dataset collected from a campus network.
The results show that while the correlation requires monitoring of custom event and flow features, it remains feasible even when using encryption protocols designed for the near future.
\end{abstract}





\section{Introduction}

Situational awareness is critical to cybersecurity in web service management as in any other computing environment.
In order to achieve a sufficient level of situational awareness, up-to-date and accurate data on what is happening are necessary.
The network flow monitoring and host-based monitoring provide orthogonal views of what is happening in the environment, but to the best of our knowledge, their monitoring data are often analyzed and evaluated separately.
With this motivation, we propose a novel approach to security monitoring using the correlation of data obtained by network flow monitoring and host-based monitoring. 

Network flow monitoring is a widely used approach to ensure a network is stable and secure~\cite{jirsik2020cyber}. 
However, it relies on deep packet inspection enrichment, hampered by currently common end-to-end encryption.
Encrypted web traffic, represented prevalently by the HTTP over TLS (HTTPS)~\cite{hu2021large}, can still be analyzed, but analysis of encrypted network data is inaccurate and costly~\cite{velan2015survey, papadogiannaki2021survey}.
The monitoring of host-based data is another well-known approach to collect and analyze network data in the form of events. 
It is not affected by end-to-end encryption; however, it requires continuous maintenance of agents and relies on accurate asset management.

The correlation of their monitoring data provides advantages for both monitoring approaches.
For network flow monitoring, the events provide metadata currently lost in encryption.
For host-based monitoring, the flows provide a consistency check; if events do not correlate with flows, it might point to a tampering attempt on a compromised server or a new web server added out of the scope of asset management.

In this paper, we investigate the event-flow correlation of HTTPS flows to web server-based events.
In particular, we seek to answer the following research questions:

\begin{enumerate}
    \item \textit{How accurately can be events recorded on a web server correlated to the network flows that caused them?}
    \item \textit{What impact will future web traffic encryption technologies have on the accuracy of the correlation process?}
\end{enumerate}

To answer the first question, we propose an event-flow correlation method for HTTP over TLS 1.2 protocol and perform it on a current dataset captured on a large campus network.
To answer the second question, we identify the limitations introduced by the new TLS 1.3 extensions and the QUIC protocol, modify the correlation method to accommodate them, and measure their impact on the correlation results.
Our results show that the event-flow correlation provides feasible results for the current HTTP over TLS 1.2 protocol, that it remains viable for TLS 1.3, and that it will cope with HTTP over QUIC if implemented in its currently drafted form~\cite{rfc-draft-http3}.


\section{Background and Related Work} \label{section-re-work}

In this section, we first set the background of our research and define the basic terms used throughout this paper.
Then we provide an overview of the state-of-the-art in encrypted traffic analysis and monitoring data correlation.

\subsection{Background}

When correlating the data obtained by network and host-based HTTPS monitoring, we work with their primary outputs: IP flows and events, respectively.

The IP flow is defined in RFC 5470: “[flow is] \textit{a set of IP packets passing an observation point in the network during a certain time interval.}”. The IPFIX protocol then defines their collecting and exporting process and specifies the number and format of the flow features~\cite{rfc5470flow, rfc7011ipfix, rfc7012features}.
It should be noted that in this research, we consider a network flow bidirectional. The bidirectional flow consolidates data transmissions in both directions from source to destination and back, as defined in RFC 5103~\cite{rfc5103biflow}. Furthermore, we refer to the flow source as the client and the destination as the server. In HTTPS web traffic, the source represents the client exclusively, as it is the client who makes a request and initiates the data exchange.

An event of a web server log is not defined as simply as the flow because there are many standards for the format of the event and for its features.
The format of the event is determined by the web server application that creates it. 
At present, the web services are mainly provided either by the Internet Information Services (IIS) on Windows Server or the Apache on Linux~\cite{iis, apache}.
These applications use different logging standards, even though both are based on the World Wide Web Consortium's Common Logging Format (CLF)~\cite{clf}. While IIS uses its own proprietary standard~\cite{iis}, Apache uses Extended Logging Format (ELF)~\cite{elf}.
Our research focuses on web services running on Windows Server; therefore, we work with events in the IIS standard. However, the proposed approach is also valid for Apache and ELF if the data normalization process is adapted for this format.

\subsection{Related Work}

We identified two areas of research that relate to our topic.
First, we describe works that focus on analysing encrypted HTTPS traffic. They show what monitoring data can be extracted compared to unencrypted HTTP and what is being lost in encryption.
Second, we discuss the papers that focus on correlating events, flows, and other types of monitoring data. We examine the features and algorithms used and check whether they are applicable in our work.

Passive monitoring and analysis of encrypted network traffic was described in depth in surveys by Velan et al. and more recently by Papadogiannaki and Ioannidis~\cite{velan2015survey, papadogiannaki2021survey}.
The surveys imply that these techniques are impeded by network traffic encryption as they rely on deep packet inspection and features that are unavailable in encrypted traffic.
Most current approaches explicitly designed for encrypted traffic monitoring and analysis deal with the missing features by relying on statistical features, e.g., number of packets, bytes, packet inter-arrival times, and using machine learning techniques, e.g., neural networks and deep learning~\cite{barut2020tls, zhang2019stnn, lotfollahi2020deep}.

According to a survey on HTTPS traffic and services identification methods by Shbair et al., HTTPS monitoring and analysis research is focused on statistical features and machine learning~\cite{shbair2020survey}. 
A novel approach to classify TLS-encrypted traffic using a neural network and autoencoder was proposed by Yang et al.~\cite{yang2018tls}.
More fine-grained classification of services or even user actions carried out over HTTPS is also possible.
Brissaud et al. proposed an approach to classify predefined user actions over the web~\cite{brissaud2019transparent} and then extended the work for detection of predefined malicious user activity in TLS encrypted HTTPS traffic~\cite{brissaud2018passive, brissaud2020encrypted}.
Shbair et al. also proposed a machine learning method for classifying HTTPS services using statistical features of network flows~\cite{shbair2020early}.
However, statistical features are unreliable, and complex machine learning techniques like neural networks often behave as a black box, where it is impossible to see the reasoning behind a result.
On the other hand, our approach proposes to transparently match the encrypted network flows with reliable event features gathered from web servers involved in the communication.

Encrypted traffic may also be monitored actively by interception proxies.
These proxies decrypt the traffic, analyze it, and then re-encrypt it, thus reducing the problem to analysis of plain network traffic~\cite{fireeye-interception, trusted-proxy}.
However, this approach invades user privacy, requires the institution that employs it to have the authority and trust of its users, and introduces security issues of its own~\cite{shbair2020survey}.
In contrast, our approach supplements the features in encrypted network traffic on the basis of their related events captured on the web server.
The communication remains encrypted, and events with all their features are already available to the web server administrator. Any features of network traffic that are not part of the web server events remain secret.
Our approach thus provides the administrator with an insight into encrypted traffic while the users retain the privacy provided by encryption.

The research of algorithms for monitoring data correlation was described in a survey by Mirheidari et al.~\cite{mirheidari2013alert}.
The survey focuses on the correlation of alerts, but it is still relevant for our research, as both an event and a flow may be abstracted as alerts from different sensors for a single occurrence.
Further, Haas et al. proposed the Zeek-osquery platform for correlating network flows with the originating processes and  users~\cite{haas2020zeek}.
Henderson et al. proposed a time-based correlation algorithm and confirmed that this approach is viable by testing it with real network data~\cite{henderson2019correlation}. 
They also discussed the limitations of the event-flow correlation. However, they investigated the correlation solely from the malicious event standpoint and did not consider network flow features aside from its start time, end time, and source. Furthermore, previous works considered the captured times of all correlated occurrences as synchronized and accurate, which is usually not the case when correlating data from devices across the network, as described by Brilingaite et al.~\cite{brilingaite2018time}.
Finally, our previous work described the correlation of events and flows of the DNS protocol using a method based on common features and a time-window to compensate inaccuracies in event and flow capture times~\cite{spacek2021enriching}.

\section{Methodology} \label{section-meth}

Our goal was to design and evaluate a method for correlating HTTPS events and flows. We chose the following approach.
First, we analyzed samples of HTTPS network flows and web server events and identified their common features.
Based on the common features, we designed the \textit{all-params} method to correlate HTTPS events and flows encrypted by TLS 1.2 protocol.
Then we designed three variants of the \textit{all-params} method to address new encryption protocols and various web server setups. The \textit{no-sni} variant is intended for flows encrypted by TLS 1.3 and QUIC, and the \textit{no-port} variant for events from web servers unable to log client port used for communication. The last \textit{no-port-sni} variant combines the adjustments from both the previous ones.
Then we collected a dataset containing HTTPS events and flows and labeled it using the \textit{all-params} method and a correlation time-window. We discuss the limitations of this approach in Section~\ref{section-lessons}.
Finally, we evaluated the \textit{no-sni}, \textit{no-port}, and \textit{no-port-sni} variants of the correlation method.

\begin{figure}
  \centering
  \includegraphics{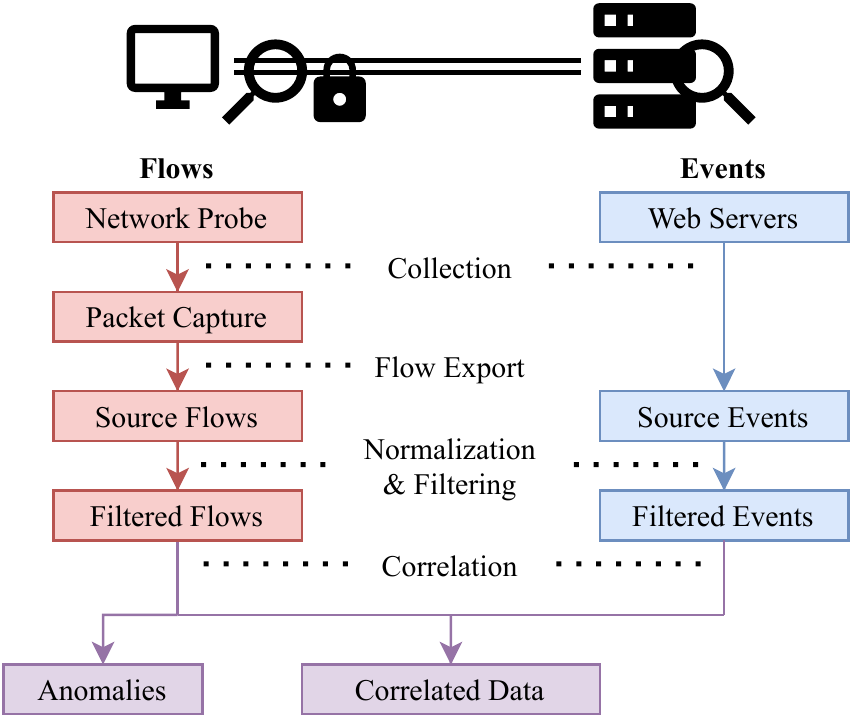}
  \caption{The process of collection, preprocessing, and correlation of HTTPS events and flows.}
  \label{fig:data-phases}
  \vspace{-5mm}
\end{figure}

The event and flow data originate from a network environment containing IIS web servers offering publicly available websites.
Clients from outside of the network communicate with the web servers using the encrypted HTTPS protocol.
Web servers log interactions with clients and thus serve as the source of the events. 
The traffic containing client requests and server responses is captured by a network traffic probe located on a link in front of the web servers.
The event and flow data captured in this environment go through several stages of processing, as shown in Figure~\ref{fig:data-phases}.
The network data is collected in a PCAP file and then transformed into flows, while the events are collected directly from the web servers.
Both types of data go through a normalization and filtering process to transform all the features into a uniform format and filter any errors and monitoring anomalies.
Finally, the events and flows are correlated and divided between correlated data and anomalies based on the correlation results.

\section{HTTPS Event-Flow Correlation} \label{section-corr-theory}

This section details the theoretical foundations of HTTPS event-flow correlation; it deals with the enumeration of common features, the definition of correlation methods, and the correctness analysis of the correlation results.

\subsection{Common Features of HTTPS Events and Flows}

\begin{table*}[ht!]
    \centering
    \caption{The features common to HTTPS events and flows and their availability in network traffic for different encryption protocols.}
    \vspace{-2mm}
    \begin{tabular}{ |l|l|l|c|c|c|c| }
        \hline
		\multicolumn{2}{|c|}{\text{Feature}} & \multicolumn{1}{c|}{\text{Description}} & \multicolumn{4}{c|}{\text{HTTP}} \\
		\multicolumn{1}{|c}{\text{Event}} & \multicolumn{1}{c|}{\text{Flow}} &  & \multicolumn{1}{c}{\text{Plain}} & \multicolumn{1}{c}{\text{TLS 1.2}} & \multicolumn{1}{c}{\text{TLS 1.3}} & \multicolumn{1}{c|}{\text{QUIC}} \\
		\hline
		time-generated & [START\_NSEC, END\_NSEC] & The time/interval of occurrence in milliseconds & \checkmark & \checkmark & \checkmark & \checkmark \\
		\hline
        s-ip & L3\_IPV4\_DST & The IP address of the logging web server & \checkmark & \checkmark & \checkmark & \checkmark \\
        \hline
        s-port & L4\_PORT\_DST & The server port number that is configured for the service & \checkmark & \checkmark & \checkmark & \checkmark \\
        \hline
        c-ip & L3\_IPV4\_SRC & The IP address of the client that made the request & \checkmark & \checkmark & \checkmark & \checkmark \\
        \hline
        c-port & L4\_PORT\_SRC & The port of the client that made the request & \checkmark & \checkmark & \checkmark & \checkmark \\
        \hline
        sc-bytes & BYTES\_B & The number of bytes that the server sent & \checkmark & \checkmark & \checkmark & \checkmark \\
        \hline
        cs-bytes & BYTES\_A & The number of bytes that the server received & \checkmark & \checkmark & \checkmark & \checkmark \\
        \hline
        cs-host & HTTP\_REQUEST\_HOST & The server name identifier (SNI) & \checkmark & \checkmark & - & - \\
        \hline
        cs-uri-stem & HTTP\_REQUEST\_URL & The resource targeted by the request & \checkmark & - & - & - \\
        \hline
        cs-user-agent & HTTP\_USER\_AGENT & The browser type that the client used & \checkmark & - & - & - \\
        \hline
    \end{tabular}
    \label{cor-features}
    \vspace{-6mm}
\end{table*}

The events and flows are different data types, but they are connected through common features.
When these features contain the same values in an event and a flow, then the event and flow may be related.
All the common correlation features for HTTPS events and flows that we identified are summarized in Table~\ref{cor-features} under names that correspond with IIS logging and the IPFIX standard.
The table also displays the availability of the features in flows encrypted by different protocols.

Most common features are present in IIS and other web servers' event logs by default. However, the client port feature is optional, and capturing it requires server configuration changes. Nonetheless, collecting the client port is important for event-flow correlation because it significantly influences its accuracy.
In the IIS environment, client port logging can be set from IIS version 8.5.
However, older servers running IIS 7.5 may still be encountered where this setting is not present.
We evaluated the correlation algorithm without using the client port to test the correlation in environments where it is only possible to rely upon basic features.

Our correlation method includes all the identified common features, with two exceptions.
The volume of transferred data, measured separately for the \textit{client-server} and \textit{server-client} direction, cannot be directly used for correlation without thorough analysis.
Such analysis falls out of the scope of this paper, so the data volume features were omitted.

\subsection{HTTPS Event-Flow Correlation Method}\label{sec-cor-methods}

We designed a correlation method to identify relations between the HTTPS events and flows.
This method is referred to as \textit{all-params} and the correlation process is described by the Algorithm~\ref{alg-all-params}.
The method is based on the common features we identified for HTTP over TLS 1.2 and a given correlation time-window. 

\vspace{-1mm}
\begin{algorithm}
\caption{Correlation algorithm \textit{all-params}}
\label{alg-all-params}
\begin{algorithmic}
    \FORALL{$flow \in flows$}
        \FORALL{$event \in events$}
            \IF{$flow.L3\_IPV4\_DST$ = $event.s\_ip$ \AND\newline
                $flow.L3\_IPV4\_SRC = event.c\_ip$ \AND\newline 
                $flow.L4\_PORT\_DST = event.s\_port$ \AND\newline
                $flow.L4\_PORT\_SRC = event.c\_port$ \AND\newline
                $flow.HTTP\_REQUEST\_HOST = event.cs\_host$ \AND\newline
                $flow.START\_NSEC - earliness \leq event.time\_generated$ \AND\newline
                $flow.END\_NSEC + lateness \geq event.time\_generated$}
                \STATE match $flow$ with $event$
            \ENDIF
        \ENDFOR
    \ENDFOR
\end{algorithmic}
\end{algorithm}
\vspace{-4mm}

Correlating events with flows exclusively when they occurred between the start and end of the flow performs poorly in the real environment due to latency, jitter, low event timestamp precision, and time synchronization drift in the network.
Consequently, usage of the correlation time-window is necessary. 
The time-window is an interval in seconds and is defined by two features -- the \textit{earliness} and \textit{lateness}. The \textit{earliness} (\textit{lateness}) is the lower (upper) bound of the time-window, and it specifies the time interval by which an event may precede (follow) a flow to be still considered related.

For our research, we used the \textit{all-params} correlation method to establish the ground truth in our dataset and determined the correlation time-window experimentally.
This process can be reused, but the ground truth and time-window need to be reestablished for any new web server environment or dataset.

We correlated the events and flows in our dataset with different time-window sizes and monitored the correct and anomalous relations counts.
To establish the ground truth, we chose the time-window that maximized the number of correct relationships and minimized the number of errors.

The \textit{all-params} correlation method is fully applicable for the TLS 1.2 protocol. The TLS 1.3 is also compatible, but only when the SNI feature is not encrypted. SNI encryption for TLS 1.3 is described in the rfc 8744~\cite{rfc7844sni}.
To extend the usability of the correlation method in environments unable to log client port and for protocols that encrypt the SNI, we designed three variants that use reduced sets of common features.
All variants of the correlation method along with used features are summarized in Table~\ref{cor-methods}.

\begin{table}[ht!]
    \vspace{-2mm}
    \centering
    \caption{Variants of the HTTPS event-flow correlation method.}
    \vspace{-2mm}
    \begin{tabular}{ |l|c|c|c|c|c| }
        \hline
        Feature & all-params &  no-sni & no-port & no-port-sni\\
        \hline
		Time of occurrence & \checkmark & \checkmark & \checkmark & \checkmark \\
		\hline
        Server IP & \checkmark & \checkmark & \checkmark & \checkmark \\
        \hline
        Server port & \checkmark & \checkmark & \checkmark & \checkmark \\
        \hline
        Client IP & \checkmark & \checkmark & \checkmark & \checkmark \\
        \hline
        Client port & \checkmark & \checkmark & - & - \\
        \hline
        SNI & \checkmark & - & \checkmark & - \\
        \hline
    \end{tabular}
    \label{cor-methods}
    \vspace{-3mm}
\end{table}

\subsection{HTTPS Event-Flow Relationships}

The event-flow correlation forms relationships between events and flows, but we cannot automatically consider all relationships created this way to be correct.
We focus on the cardinality of a relationship created by the event-flow correlation to determine its correctness.
Based on cardinality, we define correct and anomalous relations and analyze the causes of such anomalies. We start from the following two assumptions:

\begin{enumerate}
	\item \textit{Each HTTPS flow triggered at least one event at the web server.}
    \item \textit{Each event captured on a web server was caused by exactly one HTTPS flow.}
\end{enumerate}

The cardinality options for the relationship between events and flows are shown in Table~\ref{cor-states}.
The error \textit{ERR1} indicates flows and events which break the first rule, as no counterpart for them was found in correlation.
The error \textit{ERR2} includes events that break the second rule as they have been related to multiple flows at once. 

\begin{table}[ht!]
\vspace{-4mm}
    \centering
    \caption{Cardinality of all possible HTTPS event-flow relations ($m, n > 1$).}
    \vspace{-2mm}
    \begin{tabular}{ |c|c|c|l| }
        \hline
        Events & Flows & Correctness & \multicolumn{1}{c|}{\text{Description}} \\
        \hline
        1 & 0 & ERR1 & An unmatched event \\ 
        \hline
        0 & 1 & ERR1 & An unmatched flow \\
        \hline
        1 & 1 & OK & An event matched with a flow \\
        \hline
        m & 1 & OK & Events matched with a flow \\
        \hline
        1 & n & ERR2 & An event matched with multiple flows\\
        \hline
        m & n & ERR2 & Events matched with multiple flows\\
        \hline
    \end{tabular}
    \label{cor-states}
\end{table}
\vspace{-2mm}

The \textit{ERR1} correlation error can be caused during the data collection or correlation process.
Events or flows may be missing from the dataset due to a monitoring outage, and incomplete flows and events are discarded during normalization.
During the correlation process, an error of this type is caused by a too strict time-window.
Some relationships will not be established if the time-window is too small because the flow and event are too far apart.
Flows and events that remain uncorrelated are further referred to as single events and single flows. 

The \textit{ERR2} error can be caused only during correlation and applies only to events; such a relation is considered correct for the flow.
The error is caused by the inability to correctly assign an event to a flow when they match in all correlation features, and it occurs if such identical events and flows appear closer apart than is the correlation time-window.
For example, this can be caused by a web crawler repeatedly requesting the same resource from the server.
We refer to such events associated with multiple flows as polygamous events.

\section{Dataset} \label{section-dataset}

The dataset was acquired from a university network where eight web servers provide more than 800 websites.
The flows were collected with the help of a network probe situated in front of the web servers.
Events were sent from the web servers to a central log server, which collected and stored them.
All devices were time-synchronized using the Network Time Protocol (NTP).
Data collection took place for seven days, from the 30th of July to the 6th of August 2021. 

The events were collected from all web servers connected to the network.
The servers run Windows Server 2016 and therefore log using the IIS version 8.5 standard.
We used the basic IIS logging settings with three optional features enabled.
The key feature was the client port, and we also enabled logging of the volume of transferred data for both directions -- client-server and server-client. 

Network communication was captured on a probe measuring the traffic to and from the ISP of the university. Full packet capture of the web traffic on ports 80 and 443 was performed to retain as much information as possible. The first step was to reorder the packets in the PCAP file according to packet timestamps. This step was necessary because each direction of the traffic was captured on a different network interface, causing delays to be introduced by various buffers. Then we used Flowmon exporter\cite{FlowmonNetworks--Flowmon} software to generate flow records from the traffic. The exporter is able to provide SNI from TLS connections as well as properties from unencrypted HTTP headers; see Table~\ref{cor-features} for the list of primary exported features.

Finally, normalization and filtering operations were performed on the dataset.
The normalization process ensured that all common features, e.g., timestamps, were in a uniform format.
The filtering process ensured that the dataset did not contain entries with malformed or missing common features.
Both processes are described in detail by the code of our open-source software~\cite{cor-script-review}.
The resulting dataset after normalization and filtering contains 5,805,844 events and 2,836,952 flows.
		
\section{Evaluation} \label{section-eval}
The first part of this section describes the process of finding the optimal correlation time-window, which is a key part in establishing the ground truth of event-flow relations in the evaluation dataset.
The second part shows evaluation of the \textit{no-sni}, \textit{no-port}, and \textit{no-port-sni} event-flow correlation variants described in Section~\ref{sec-cor-methods}.

\subsection{Time-Window Measurement}
 
To compute the optimal time-window, we used a weighted method minimizing the number of the erroneous \textit{ERR1} and \textit{ERR2} correlation results.
The weights of \textit{ERR1} and \textit{ERR2} errors were set to 1 and 2, respectively. 
The reason for such an imbalanced weight distribution is that in terms of significance, the \textit{ERR2} is a correlation error with the same impact as \textit{ERR1}. However, it only occurs for events because it is considered a correct state for flows (see Table~\ref{cor-states}). 
Thus, with the same weights, the resulting time-window would favor a lower number of relationships with error \textit{ERR1} over \textit{ERR2}.
Then we performed correlations with 36 time-windows combining all values of earliness and lateness from an interval $<0,1..5>$ seconds.
We consider five seconds to be a sufficient interval to cope with latency, jitter, and time synchronization drift in the environment. 
If the distance between an event and a flow is greater, it is an anomaly that ranks such an event-flow pair among the uncorrelated data even if their other features match. 

\begin{table*}[ht!]
    \centering
    \caption{The effect of different time-window sizes on the results of \textit{all-params} correlation.\\ The time-window format is (earliness, lateness).}
    \vspace{-2mm}
    \begin{tabular}{ |l|r|r|r|r|r|r|r|r|r|r|r|r|r|r| }
        \hline
        Time-Window Size & \multicolumn{1}{c|}{\text{(5, 0)}} & \multicolumn{1}{c|}{\text{(3, 0)}} & \multicolumn{1}{c|}{\text{(2, 0)}} & \multicolumn{1}{c|}{\text{(1, 0)}} & \multicolumn{1}{c|}{\text{(0, 0)}} & \multicolumn{1}{c|}{\text{(0, 1)}} & \multicolumn{1}{c|}{\text{(0, 2)}} & \multicolumn{1}{c|}{\text{(0, 3)}} & \multicolumn{1}{c|}{\text{(0, 5)}} & \multicolumn{1}{c|}{\text{(NA, NA)}}\\
        \hline
        Single Flows & 380028 & 380036 & 391270 & 966928 & 2242420 & 2242356 & 2242286 & 2242241 & 2242152 & 376012 \\
        \hline
        Correlated Flows & 2456924 & 2456916 & 2445682 & 1870024 & 594532 & 594596 & 594666 & 594711 & 594800 & 2460940 \\
        \hline
        Single Events & 193176 & 193176 & 208258 & 1173216 & 3431247 & 3431120 & 3430963 & 3430838 & 3430597 & 95557 \\
        \hline
        Correlated Events & 5612552 & 5612595 & 5597527 & 4632605 & 2374597 & 2374713 & 2374846 & 2374951 & 2375141 & 5089360 \\
        \hline
        Polygamous Events & 116 & 73 & 59 & 23 & 0 & 11 & 35 & 55 & 106 & 620927 \\
        \hline
    \end{tabular}
    \label{time-windows}
    \vspace{-5mm}
\end{table*}

We list the correlation results for significant time-windows in Table~\ref{time-windows} where we monitor the number of successfully correlated events and flows, as well as the count of \textit{ERR1} and \textit{ERR2} correlation errors.
The time-window (0, 0) corresponds to a correlation with no tolerance interval. The results of this correlation show that a time-window is indeed needed because only 45,8\% of events and 26,2\% of flows could be correlated. 
The time-window (NA, NA) corresponds to a correlation over the whole dataset regardless of the time of occurrence. It represents the maximum possible number of related events and flows that can be found in the dataset. However, in this case, even events and flows divided by hours will be considered related, which is not a reasonable assumption.

The time-windows (5, 0) -- (0, 5) in Table~\ref{time-windows} illustrate the effect of changing earliness and lateness on correlation results.
In our environment, rising earliness to three seconds increased the number of successfully paired events and flows by nearly two million.
The lateness had a significantly lower impact on the results, finding only hundreds of new relations.
The \textit{all-params} correlation method showed the lowest number of errors in the time-window (3, 0), where 86,6\% of flows and 96,7\% of events were correlated.
Consequently, the relationships between events and flows identified in this time-window are considered the ground truth.

\subsection{Correlation Method Evaluation}
The first variant of the \textit{all-params} correlation method is the \textit{no-sni} method. 
It omits the SNI from the correlation features and is intended for network traffic in which this parameter is unavailable, e.g., TLS 1.3 and QUIC network flows. 
Such a weakening of the correlation rules results in less accurate correlation and a slight deterioration in the monitored metrics because the method also creates relationships between flows and events that differ only in the SNI feature.
According to the results in Table~\ref{cor-eval}, there are not many such relationships in the dataset, because the deterioration in \textit{accuracy} and \textit{recall} is only slight. 
The \textit{precision} remains unchanged because removing a correlation parameter and thus softening the correlation rules cannot result in rejection of a relationship that has been identified in the ground truth. 
With an \textit{F1-score} of 99.35\%, this method is relevant for correlating the event and the HTTPS flow.
Consequently, the event-flow correlation is viable even if the SNI is encrypted in future versions of HTTPS.

The second variant is the \textit{no-port} method.
It is intended mainly for environments that do not allow monitoring of client ports in web server events.
However, as the results for this method show in Table~\ref{cor-eval}, the client port is a key feature with a strong impact on correlation results.
When it is unavailable, the correlation is identifying a huge number of false-positive relations resulting in a \textit{precision} of only 40.55\%.
While the \textit{accuracy} and \textit{recall} remain close to one, low \textit{precision} renders this method unusable with the \textit{F1-score} of 57.70\%.
Consequently, event-flow correlation in environments not providing the client port in HTTPS communication and web server events does not provide satisfying results.

The third variant is the \textit{no-port-sni} method.
It is a combination of the \textit{no-sni} and \textit{no-port} methods for the environment where only the basic event and flow features are available.
Omitting the client port from correlation features impacts this method more than omitting the SNI, so the results in Table~\ref{cor-eval} are similar to the \textit{no-port} method.
The \textit{precision} is slightly lower, reaching only 35.55\%, and the \textit{F1-score} falls to 52.45\%.
When compared with the \textit{no-port} method, we can see that the SNI has only a marginal effect on the correlation performance. 
We conclude that event-flow correlation based on a feature set not including the client port is not feasible.

\begin{table}[ht!]
    \vspace{-2mm}
    \centering
    \caption{Evaluation of correlation methods.}
    \vspace{-2mm}
    \begin{tabular}{ |l|c|c|c|c| }
        \hline
         & all-params & no-sni & no-port & no-port-sni \\
        \hline
        Accuracy & 1.0000 & 0.9999 &  0.9999 & 0.9999 \\ 
        \hline
        Precision & 1.0000 & 0.9999 &  0.4055 & 0.3555 \\
        \hline
        Recall & 1.0000 & 1.0000 & 1.0000 & 1.0000 \\
        \hline
        F1-Score & 1.0000 & 0.9999 & 0.5770 &  0.5245\\
        \hline
    \end{tabular}
    \label{cor-eval}
    \vspace{-3mm}
\end{table}

\section{Lessons Learned} \label{section-lessons}

While performing event-flow correlation in our environment, we found that the web servers receive traffic for which we have no logged events and vice-versa.
Even when using the least limited \textit{all-params} correlation neglecting the time of occurrence -- time-window (NA, NA), 13.25\% of flows and 1.65\% of events remain uncorrelated.
In this section, we describe the factors that impacted the correlation results and the measures we took to suppress them.

The university web servers from which we collected the events are administratively fragmented.
Such fragmentation causes inconsistencies in the configuration of web servers, which makes it difficult to ensure a uniform collection of events.
Older web sites using IIS 7.5 logging caused issues because it was not possible to log either the client port or the original client IP address if the site was behind a reverse proxy.
We filtered events from these web sites from the dataset, as they could not possibly contain enough features to correlate with the flows.
However, the lack of complete knowledge of web server configuration made it impossible to set the network flow filter to exactly match the event filter.
We believe that the relaxed flow filter caused a higher percentage of single flows.

The flow exporter settings also influence the correlation results~\cite{velan2020impact}. The goal of flow exporters is to create flow records that represent connections as they appear on the network. However, since it is not feasible to keep the exact state of every connection to determine its termination, simplified conditions are applied to recognize when the flow records should end. In addition, long connections are split after an active timeout to accommodate timely reporting of the ongoing traffic.

In this paper, we have used a long active timeout to avoid generating multiple flow records for long connections since that would harm the correlation with web traffic logs. 
However, when multiple connections reuse the same IP addresses, protocol, and ports, those connections are added into a single flow record. This has a negative impact on the quality of data since a connection to flow record correspondence is impaired. Moreover, HTTP hostname and TLS SNI are extracted only from the first connection. Therefore, when a subsequent connection has different values, the corresponding log event cannot be correlated.

To mitigate this, we used a technique based on connection establishment itself. When a second SYN packet is observed for a flow record in a single direction, that flow record is immediately terminated, and a new one is established. 
This flow termination method can be easily applied to real-time traffic monitoring as well. It would allow keeping longer inactive timeouts for TCP connections to prevent unnecessary split of flow records while preventing the unwanted combination of several connections into a single flow record. The code for our plugin can be found at~\cite{cor-script-review}.

\section{Conclusion} \label{section-conclusion}

We proposed and evaluated a method for correlating web server events and network flows of the HTTPS protocol.
The event-flow correlation method successfully identified relations between events and flows encrypted by the TLS 1.2 protocol.
Performance of the correlation method's variant designed with a reduced set of common features to match flows encrypted by TLS 1.3 and QUIC also proved satisfactory.
The variant designed for web server environments without the ability to monitor client ports performed poorly, and the client port proved a key feature in HTTPS event-flow correlation.
Despite the fact, we see the event-flow correlation as a promising method of monitoring encrypted HTTPS traffic, and we provide all the code developed during our research as open-source~\cite{cor-script-review}.

%

\section*{Acknowledgment}

This research was supported by the CONCORDIA project that has received funding from the European Union’s Horizon 2020 research and innovation programme under the grant agreement No. 830927 and by the ERDF project “CyberSecurity, CyberCrime and Critical Information Infrastructures Center of Excellence” (No.CZ.02.1.01/0.0/0.0/16019/0000822).

\bibliographystyle{IEEEtran}
\bibliography{references.bib}

\begin{thebibliography}{10}
\providecommand{\url}[1]{#1}
\csname url@samestyle\endcsname
\providecommand{\newblock}{\relax}
\providecommand{\bibinfo}[2]{#2}
\providecommand{\BIBentrySTDinterwordspacing}{\spaceskip=0pt\relax}
\providecommand{\BIBentryALTinterwordstretchfactor}{4}
\providecommand{\BIBentryALTinterwordspacing}{\spaceskip=\fontdimen2\font plus
\BIBentryALTinterwordstretchfactor\fontdimen3\font minus
  \fontdimen4\font\relax}
\providecommand{\BIBforeignlanguage}[2]{{%
\expandafter\ifx\csname l@#1\endcsname\relax
\typeout{** WARNING: IEEEtran.bst: No hyphenation pattern has been}%
\typeout{** loaded for the language `#1'. Using the pattern for}%
\typeout{** the default language instead.}%
\else
\language=\csname l@#1\endcsname
\fi
#2}}
\providecommand{\BIBdecl}{\relax}
\BIBdecl

\bibitem{jirsik2020cyber}
T.~Jirsik and P.~Celeda, ``Cyber situation awareness via ip flow monitoring,''
  in \emph{NOMS 2020-2020 IEEE/IFIP Network Operations and Management
  Symposium}.\hskip 1em plus 0.5em minus 0.4em\relax IEEE, 2020, pp. 1--6.

\bibitem{hu2021large}
Q.~Hu, M.~R. Asghar, and N.~Brownlee, ``{A Large-Scale Analysis of HTTPS
  Deployments: Challenges, Solutions, and Recommendations},'' \emph{Journal of
  Computer Security}, no. Preprint, pp. 1--26, 2021.

\bibitem{velan2015survey}
P.~Velan, M.~{\v{C}}erm{\'a}k, P.~{\v{C}}eleda, and M.~Dra{\v{s}}ar, ``{A
  Survey of Methods for Encrypted Traffic Classification and Analysis},''
  \emph{International Journal of Network Management}, vol.~25, no.~5, pp.
  355--374, 2015.

\bibitem{papadogiannaki2021survey}
E.~Papadogiannaki and S.~Ioannidis, ``{A Survey on Encrypted Network Traffic
  Analysis Applications, Techniques, and Countermeasures},'' \emph{ACM
  Computing Surveys (CSUR)}, vol.~54, no.~6, pp. 1--35, 2021.

\bibitem{rfc-draft-http3}
\BIBentryALTinterwordspacing
M.~Bishop, ``{Hypertext Transfer Protocol Version 3 (HTTP/3)},'' Internet
  Engineering Task Force, Feb. 2021. [Online]. Available:
  \url{https://datatracker.ietf.org/doc/html/draft-ietf-quic-http}
\BIBentrySTDinterwordspacing

\bibitem{rfc5470flow}
\BIBentryALTinterwordspacing
C.~Huitema, ``{Architecture for IP Flow Information Export},'' Internet
  Engineering Task Force, Mar. 2009. [Online]. Available:
  \url{https://datatracker.ietf.org/doc/html/rfc5470}
\BIBentrySTDinterwordspacing

\bibitem{rfc7011ipfix}
\BIBentryALTinterwordspacing
B.~Claise, B.~Trammel, and P.~Aitken, ``{Specification of the IP Flow
  Information Export (IPFIX) Protocol for the Exchange of IP Traffic Flow
  Information},'' Internet Engineering Task Force, Sep. 2013. [Online].
  Available: \url{https://tools.ietf.org/html/rfc7011}
\BIBentrySTDinterwordspacing

\bibitem{rfc7012features}
\BIBentryALTinterwordspacing
B.~Trammel and E.~Boschi, ``{Information Model for IP Flow Information Export
  (IPFIX)},'' Internet Assigned Numbers Authority, Sep. 2008. [Online].
  Available: \url{https://www.iana.org/assignments/ipfix/ipfix.xhtml}
\BIBentrySTDinterwordspacing

\bibitem{rfc5103biflow}
\BIBentryALTinterwordspacing
------, ``{Bidirectional Flow Export Using IP Flow Information Export
  (IPFIX)},'' Internet Engineering Task Force, Jan. 2008. [Online]. Available:
  \url{https://tools.ietf.org/html/rfc5103}
\BIBentrySTDinterwordspacing

\bibitem{iis}
\BIBentryALTinterwordspacing
{Microsoft}, ``{Internet Information Services}.'' [Online]. Available:
  \url{https://www.iis.net/}
\BIBentrySTDinterwordspacing

\bibitem{apache}
\BIBentryALTinterwordspacing
{Apache Software Foundation}, ``{Apache HTTP Server Project}.'' [Online].
  Available: \url{https://httpd.apache.org/}
\BIBentrySTDinterwordspacing

\bibitem{clf}
\BIBentryALTinterwordspacing
W3C, ``{Logging Control In W3C httpd},'' World Wide Web Consortium (W3C).
  [Online]. Available: \url{https://www.w3.org/Daemon/User/Config/Logging.html}
\BIBentrySTDinterwordspacing

\bibitem{elf}
\BIBentryALTinterwordspacing
P.~M. Hallam-Baker and B.~Behlendorf, ``{Extended Log File Format},'' World
  Wide Web Consortium (W3C). [Online]. Available:
  \url{https://www.w3.org/TR/WD-logfile.html}
\BIBentrySTDinterwordspacing

\bibitem{barut2020tls}
O.~Barut, R.~Zhu, Y.~Luo, and T.~Zhang, ``{TLS Encrypted Application
  Classification Using Machine Learning with Flow Feature Engineering},'' in
  \emph{2020 the 10th International Conference on Communication and Network
  Security}, 2020, pp. 32--41.

\bibitem{zhang2019stnn}
Y.~Zhang, S.~Zhao, J.~Zhang, X.~Ma, and F.~Huang, ``{STNN: A Novel TLS/SSL
  Encrypted Traffic Classification System Based on Stereo Transform Neural
  Network},'' in \emph{2019 IEEE 25th International Conference on Parallel and
  Distributed Systems (ICPADS)}.\hskip 1em plus 0.5em minus 0.4em\relax IEEE,
  2019, pp. 907--910.

\bibitem{lotfollahi2020deep}
M.~Lotfollahi, M.~J. Siavoshani, R.~S.~H. Zade, and M.~Saberian, ``{Deep
  Packet: A Novel Approach for Encrypted Traffic Classification Using Deep
  Learning},'' \emph{Soft Computing}, vol.~24, no.~3, pp. 1999--2012, 2020.

\bibitem{shbair2020survey}
W.~M. Shbair, T.~Cholez, J.~Fran{\c{c}}ois, and I.~Chrisment, ``{A Survey of
  HTTPS Traffic and Services Identification Approaches},'' \emph{arXiv preprint
  arXiv:2008.08339}, 2020.

\bibitem{yang2018tls}
Y.~Yang, C.~Kang, G.~Gou, Z.~Li, and G.~Xiong, ``{TLS/SSL Encrypted Traffic
  Classification with Autoencoder and Convolutional Neural Network},'' in
  \emph{2018 IEEE 20th International Conference on High Performance Computing
  and Communications; IEEE 16th International Conference on Smart City; IEEE
  4th International Conference on Data Science and Systems
  (HPCC/SmartCity/DSS)}.\hskip 1em plus 0.5em minus 0.4em\relax IEEE, 2018, pp.
  362--369.

\bibitem{brissaud2019transparent}
P.-O. Brissaud, J.~Franc{\c{c}}is, I.~Chrisment, T.~Cholez, and O.~Bettan,
  ``{Transparent and Service-Agnostic Monitoring of Encrypted Web Traffic},''
  \emph{IEEE Transactions on Network and Service Management}, vol.~16, no.~3,
  pp. 842--856, 2019.

\bibitem{brissaud2018passive}
P.-O. Brissaud, J.~Fran{\c{c}}ois, I.~Chrisment, T.~Cholez, and O.~Bettan,
  ``{Passive Monitoring of HTTPS Service Use},'' in \emph{2018 14th
  International Conference on Network and Service Management (CNSM)}.\hskip 1em
  plus 0.5em minus 0.4em\relax IEEE, 2018, pp. 219--225.

\bibitem{brissaud2020encrypted}
------, ``{Encrypted HTTP/2 Traffic Monitoring: Standing the Test of Time and
  Space},'' in \emph{2020 IEEE International Workshop on Information Forensics
  and Security (WIFS)}.\hskip 1em plus 0.5em minus 0.4em\relax IEEE, 2020, pp.
  1--6.

\bibitem{shbair2020early}
W.~M. Shbair, T.~Cholez, J.~Fran{\c{c}}ois, and I.~Chrisment, ``{Early
  Identification of Services in HTTPS Traffic},'' \emph{arXiv preprint
  arXiv:2008.08350}, 2020.

\bibitem{fireeye-interception}
\BIBentryALTinterwordspacing
{FireEye}. {FireEye SSL Intercept Appliance, Expose Attacks Hiding in SSL
  Traffic}. [Online]. Available:
  \url{https://www.greenit-solution.ch/wp-content/uploads/2016/05/ds-ssl-intercept.pdf}
\BIBentrySTDinterwordspacing

\bibitem{trusted-proxy}
\BIBentryALTinterwordspacing
{I. H. W. Group}. {Explicit trusted proxy in http/2.0}. [Online]. Available:
  \url{https://tools.ietf.org/html/draft-loreto-httpbis-trusted-proxy20-01}
\BIBentrySTDinterwordspacing

\bibitem{mirheidari2013alert}
S.~A. Mirheidari, S.~Arshad, and R.~Jalili, ``{Alert Correlation Algorithms: A
  Survey and Taxonomy},'' in \emph{International Symposium on Cyberspace Safety
  and Security}.\hskip 1em plus 0.5em minus 0.4em\relax Springer, 2013, pp.
  183--197.

\bibitem{haas2020zeek}
S.~Haas, R.~Sommer, and M.~Fischer, ``{Zeek-Osquery: Host-Network Correlation
  for Advanced Monitoring and Intrusion Detection},'' \emph{arXiv preprint
  arXiv:2002.04547}, 2020.

\bibitem{henderson2019correlation}
S.~Henderson, B.~Nicholls, and B.~Ehmann, ``{Time-Based Correlation of
  Malicious Events and Their Connections},''
  \url{https://resources.sei.cmu.edu/asset_files/Presentation/2019_017_001_539987.pdf},
  accessed: 2021-09-15.

\bibitem{brilingaite2018time}
A.~Brilingait{\.e}, L.~Bukauskas, and E.~Kutka, ``{Time-Line Alignment of Cyber
  Incidents in Heterogeneous Environments},'' in \emph{ECCWS 2018 17th European
  Conference on Cyber Warfare and Security}.\hskip 1em plus 0.5em minus
  0.4em\relax Academic Conferences and publishing ltd., 2018, p.~57.

\bibitem{spacek2021enriching}
S.~{\v{S}}pa{\v{c}}ek, D.~Tovar{\v{n}}{\'a}k, and P.~{\v{C}}eleda, ``{Enriching
  DNS Flows with Host-Based Events to Bypass Future Protocol Encryption},'' in
  \emph{IFIP International Conference on ICT Systems Security and Privacy
  Protection}.\hskip 1em plus 0.5em minus 0.4em\relax Springer, 2021, pp.
  302--316.

\bibitem{rfc7844sni}
\BIBentryALTinterwordspacing
G.~Sadasivan, N.~Brownlee, B.~Claise, and J.~Quittek, ``{Issues and
  Requirements for Server Name Identification (SNI) Encryption in TLS},''
  Internet Engineering Task Force, Jul. 2020. [Online]. Available:
  \url{https://datatracker.ietf.org/doc/html/rfc8744}
\BIBentrySTDinterwordspacing

\bibitem{FlowmonNetworks--Flowmon}
\BIBentryALTinterwordspacing
{Flowmon Networks}. {Flowmon Probe}. [Online]. Available:
  \url{https://www.flowmon.com/en/products/flowmon/probe}
\BIBentrySTDinterwordspacing

\bibitem{cor-script-review}
\BIBentryALTinterwordspacing
S.~{\v{S}}pa{\v{c}}ek, P.~Velan, P.~Čeleda, and D.~Tovarňák, ``{HTTPS
  Event-Flow Correlation Improving Situational Awareness in Encrypted Web
  Traffic - Data and Code},'' \emph{Zenodo}, Jan. 2022. [Online]. Available:
  \url{https://zenodo.org/record/5821815}
\BIBentrySTDinterwordspacing

\bibitem{velan2020impact}
P.~Velan and T.~Jirsik, ``On the impact of flow monitoring configuration,'' in
  \emph{NOMS 2020-2020 IEEE/IFIP Network Operations and Management
  Symposium}.\hskip 1em plus 0.5em minus 0.4em\relax IEEE, 2020, pp. 1--7.

\end{thebibliography}

\end{document}